\RequirePackage{fix-cm}
\documentclass[a4paper]{svjour3}                     
\usepackage{geometry}
\smartqed  
\usepackage[pdftex]{graphicx}
\usepackage{mathptmx}      
\usepackage{latexsym}
\usepackage{amsmath}
\usepackage{algorithm}
\usepackage{algpseudocode}
\usepackage{color}
\usepackage[misc]{ifsym}

\begin{document}

\title{VNE Solution for Network Differentiated QoS and Security Requirements: From the Perspective of Deep Reinforcement Learning}


\author{Chao Wang \and Ranbir Singh Batth \and Peiying Zhang \and Gagangeet Singh Aujla \and Youxiang Duan \and Lihua Ren}


\institute{C. Wang \at
              China University of Petroleum (East China), Qingdao, China. \\
              \email{wangch\_upc@qq.com} \\
              \and
              R. S. Batth (\Letter)\at
              Lovely Professional University, Phagwara, Punjab, India. \\
              \email{ranbir.21123@lpu.co.in}
              \and
              P. Zhang (\Letter) \at
              China University of Petroleum (East China), Qingdao, China. \\
              \email{zhangpeiying@upc.edu.cn} \\
              \and
              G. S. Aujla \at
              Durham University, Durham, United Kingdom. \\
              \email{gagi\_aujla82@yahoo.com and gagangeet.s.aujla@durham.ac.uk}
              \and
              Y. Duan \and L. Ren \at
              China University of Petroleum (East China), Qingdao, China. \\
              \\
              Y. Duan \\
              \email{yxduan@upc.edu.cn} \\
              \\
              L. Ren\\
              \email{renlh@upc.edu.cn}
}
\date{Received: date / Accepted: date}

\maketitle

\begin{abstract}
The rapid development and deployment of network services has brought a series of challenges to researchers. On the one hand, the needs of Internet end users/applications reflect the characteristics of travel alienation, and they pursue different perspectives of service quality. On the other hand, with the explosive growth of information in the era of big data, a lot of private information is stored in the network. End users/applications naturally start to pay attention to network security. In order to solve the requirements of differentiated quality of service (QoS) and security, this paper proposes a virtual network embedding (VNE) algorithm based on deep reinforcement learning (DRL), aiming at the CPU, bandwidth, delay and security attributes of substrate network. DRL agent is trained in the network environment constructed by the above attributes. The purpose is to deduce the mapping probability of each substrate node and map the virtual node according to this probability. Finally, the breadth first strategy (BFS) is used to map the virtual links. In the experimental stage, the algorithm based on DRL is compared with other representative algorithms in three aspects: long term average revenue, long term revenue consumption ratio and acceptance rate. The results show that the algorithm proposed in this paper has achieved good experimental results, which proves that the algorithm can be effectively applied to solve the end user/application differentiated QoS and security requirements.
\keywords{End User/Application \and Quality of Service \and Security Requirement \and Virtual Network Embedding}
\end{abstract}

\section{Introduction}

The rapid development of network technology and the blowout growth of end users/applications have brought greater opportunities and challenges to infrastructure providers (InPs) and service providers (SPs) \cite{z1,g1}. With the 5G era, many network applications are based on virtual network architecture. In this architecture, the implementation of network functions no longer depends on specific hardware facilities, but on the way of software programming to achieve flexible deployment of virtual functions \cite{n1,j4,a1}. For example, the realization of intelligent applications, such as the Internet of vehicles (IoV), intelligent medical, military unmanned aerial vehicle (UAV), cannot do without a strong virtual network infrastructure as support \cite{g7,b1,s3}. It cannot be ignored that there are a series of challenges in using network virtualization (NV) technology to provide services for these intelligent applications \cite{j1,z2}. Radio network resource management faces severe challenges, including storage, spectrum, computing resource allocation, and joint allocation of multiple resources \cite{jcx1,jcx2}. With the rapid development of communication networks, the integrated space-ground network has also become a key research object \cite{jcx3}.

On the one hand, the QoS is closely related to the needs of network end users/applications. With the growth of network end users/applications and the expansion of network services, the QoS requirements of users/applications show the characteristics of differentiation \cite{b2}. For the users of the IoV or UAVs, their primary demand is the real-time and accurate control of the vehicles or UAVs \cite{s1}. Driverless cars need to judge the road conditions and make accurate decisions in time to avoid traffic accidents \cite{g4,s4}. UAV needs timely command to strike the target accurately. So they need InPs to provide low latency network services. For the network video live users, they need to use the network to carry out live activities. This type of application requires the network to provide a large amount of bandwidth in a short period of time to ensure the smoothness of the video picture, so it puts forward a high bandwidth service demand for the InPs. In addition, it also includes low cost, low memory consumption and other different QoS requirements. Therefore, large-scale end users/applications put forward differentiated QoS requirements for InPs \cite{a2,k4,k5,k6}.

On the other hand, with the mutual penetration of network technology and daily life, people's demand of using network to store personal information is more and more intense \cite{g2,b3,k1,k2,k3}. For bank accounts, health monitoring and electronic payment, these important private information is not expected to be exposed or stolen, and this information is the target of malicious devices or malware attacks \cite{s5,s6}. Therefore, the security of network services should be considered when satisfying the differentiated QoS requirements of network end users/applications \cite{a3,a4,g3,s7}.

In the NV environment, the requirements of network end users/applications are presented in the form of virtual network requests (VNRs). The SP is responsible for sending the VNR of the network end user/application. The purpose is to hope that the InP can allocate sufficient underlying network resources to meet the demand, that is, the problem of VNE \cite{z3,g5,s8}. This paper focuses on the problem of VNE with differentiated QoS requirements and security. According to the functional requirements of different users, this paper focuses on the design of differentiated QoS VNE algorithm from three aspects of VNE cost, network bandwidth and delay. In view of the security problems exposed in the process of VNE, a reasonable security level is set for VNR and substrate network. A VNE algorithm is designed to meet the requirements of differentiated QoS and security.

In order to improve the decision-making and optimization ability of VNE algorithm, deep learning method and reinforcement learning method are applied. The deep learning method is mainly used to solve the decision-making problem in high-dimensional space. By imitating the biological neural network to establish the network model, the problem that needs to be decided is input into the neural network and finally the optimal solution of the problem can be obtained \cite{g6,n3}. The reinforcement learning method mainly emphasizes the learning and training of agents in the interaction with the environment, and realizes the optimization of decision-making by using the evaluation feedback signal. In reinforcement learning method, training set data is usually used to train the agent and the agent adjusts the action through the size of feedback signal, so as to achieve the effect of continuous optimization. Finally, the optimal result can be obtained in the test data set \cite{j2,j3,g8}. The embedded problem of virtual network has been proved to be NP-hard \cite{z4}, so deep reinforcement learning can be used to find the optimal solution for the embedded problem of virtual network to meet the differentiated QoS and security needs of network end users/applications.

The main work of this paper is as follows:

(1) Aiming at the differentiated QoS and security requirements of network end users/applications, this paper applies the representative DRL algorithm to the VNE problem. Through the efficient training results of agents, the differentiated QoS and security requirements of network end users/applications can be effectively solved.

(2) In the DRL algorithm, we extract four important attributes for the substrate network: CPU, bandwidth, delay and security. Using the custom policy network as the agent, the feature matrix is used as the input of the policy network for training. In this way, agents can be trained in a more realistic network environment, and the experimental results are also optimal. Finally, the mapping probability of each substrate node can be obtained.

(3) In order to prove the effectiveness of the algorithm, the VNE algorithm based on DRL is compared with other representative algorithms in three aspects of long term average revenue, long term revenue consumption ratio and acceptance rate. The experimental results show that the algorithm achieves good results and proves the effectiveness of the algorithm.

The rest of this paper is organized as follows: The second part describes the research status of VNE algorithms for differentiated QoS and security. The third part describes the problem of VNE and establishes the network model. The fourth part describes the implementation process of VNE algorithm based on differentiated QoS and security requirements. The fifth part introduces the setting of simulation experiment, then shows the experimental results and analyzes them. The last part summarizes the whole paper.

\section{Related Work}

\subsection{VNE Algorithms Based on Differentiated QoS}

In reference \cite{c1}, a dynamic heuristic algorithm is proposed, which focuses on receiving as many VNRs as possible instead of optimizing the QoS performance of each VNR. When the QoS of VNR is not satisfied, the algorithm will drive the re-embedding scheme of heuristic algorithm to meet the given QoS requirements. Reference \cite{c2} considers that the existing solutions only aim at the congestion control problem of single objective VNE, and proposes a multi-objective VNE solution. Aiming at energy saving, energy sensing, avoiding network congestion and other service indicators, the embedding process of virtual network is completed by combining the heuristic solution method based on SDN. In reference \cite{c3}, a dynamic network resource allocation method based on load balancing and QoS is proposed. In this paper, the author proposes a QoS based scheduling mechanism for VNRs, which can reasonably rank incoming services by calculating the priority of VNRs. At the same time, this method uses a resource allocation mechanism based on load balancing to avoid the imbalance of resource consumption. In reference \cite{c4}, an intelligent delay aware VNE scheme iVNE is proposed. This scheme focuses on the problem that the existing embedding algorithm of virtual network is not necessarily the optimal algorithm of industrial wireless network, but also lacks the QoS capacity. It provides delay guarantee for various industrial virtual networks, including static embedding process and dynamic forwarding process. Finally, the VNE algorithm achieves good load balancing ability. Reference \cite{s2} pays attention to the resource constraints and service quality issues of the IoV scenario. Based on artificial intelligence and machine learning, the author pushes cache and communication resources to the edge of smart cars, and jointly realizes the offloading of roadside units (RSU). The author uses a mixed integer nonlinear programming (MINLP) model to reduce the total network delay. The final experimental results prove that this method is effective in reducing user communication, computing, network congestion and content download delays.

\subsection{VNE Algorithms Based on Security}

In reference \cite{c5}, network function virtualization (NFV) technology is applied to the field of network security. The network services processed by virtual network security function chain may be sensitive to specific network requirements, such as maximum bandwidth or minimum delay. The author proposes a gradual security service embedding scheme, which can optimize the resource utilization and deploy the virtual security function chain efficiently according to the security requirements of a single application and the strategy of the operator. In reference \cite{c6}, a heuristic security aware VNE algorithm (SA-VNE) is proposed. The algorithm uses TOPSIS to sort the importance of the base nodes and select the most suitable base nodes. Finally, the shortest path algorithm is used to complete the link mapping process. In reference \cite{c7}, trust relationship and trust degree are introduced into the problem of VNE, and the security problems in NV environment are analyzed quantitatively. This paper proposes a trust aware security VNE algorithm, which considers the local and global importance of nodes in the mapping process, and uses the approximate ideal ranking method to sort the substrate nodes. Finally, the k-shortest path method is used to complete the link mapping. In reference \cite{c8}, the mapping process of security virtual network is modeled as a multi-objective mixed integer linear programming model, and a mapping algorithm of security virtual network based on multi-attribute comprehensive evaluation of nodes and path optimization is proposed. In this algorithm, the resource richness, security attributes and topological proximity of nodes are regarded as the criteria of node selection. In the link mapping phase, the available bandwidth and the number of path hops are used as the evaluation objects to select the mapping link. Finally the whole VNE process is completed.

From the above VNE algorithm based on differentiated QoS requirements and the security VNE algorithm, the existing VNE algorithm has done more perfect work. It cannot be ignored that they still have the following problems. First of all, in the research of VNE for QoS requirements, the author does not clearly point out which specific QoS indicators are. They regard the whole QoS as an evaluation standard, which does not reflect the characteristics of differentiation. There are no specific examples of differentiated QoS that need to be solved in the existing research, lacking of universal practical significance. Secondly, the research of security VNE algorithm only uses the traditional heuristic method. With the rapid development of intelligent learning method, it is of great significance to apply it to the problem of VNE. The intelligent learning method can allocate the network resources satisfying the security characteristics of VNRs, which has obvious advantages over the traditional heuristic method. Finally, there is no research on VNE algorithm which combines differentiated QoS requirements and security. In this paper, we will combine the differentiated QoS requirements and security issues, and apply DRL method to study the VNE algorithm.

\section{Description and Model Establishment of VNE Problem with Differentiated QoS and Security}

\subsection{Description of VNE Problem with Differentiated QoS and Security}

The different network functional requirements of network end users/applications are multiple heterogeneous VNRs. To allocate the underlying network resources reasonably for these VNRs, this process is called VNE. The embedding of virtual network can be divided into two parts: node embedding and link embedding \cite{j5,z5}.

In the node embedding stage, virtual nodes need to find substrate nodes to meet their resource requirements. In order to better solve the problem of VNE with differentiated QoS requirements, we will focus on the cost of CPU resource consumption and the impact of node delay on QoS. For the security problem, the security requirement level attribute will be set for each virtual node, and each virtual node can only be embedded in the substrate node that meets its security requirements. Therefore, CPU resource, delay and security are three important indexes in the node attribute setting.

In the stage of link embedding, virtual link needs to find a substrate link to meet its resource requirements. A virtual link can be embedded in one substrate link or multiple substrate links through path segmentation. In order to better solve the problem of VNE with differentiated QoS requirements, we will focus on the cost of bandwidth consumption and the impact of link delay on QoS. Therefore, bandwidth resource and delay are two important indexes in link attribute setting.

In the whole process of VNE, CPU, bandwidth, delay and security are taken as the starting point to solve the problem of VNE with differentiated QoS requirements and security. By setting reasonable attributes for nodes and links, a reliable solution is provided for the problem of VNE facing differentiated QoS requirements and security.

\subsection{Network Models}

The undirected weighted graph $G^V=\{N^V,L^V,A_N^V,A_L^V\}$ is used to model the virtual network. $G^V$ represents a separate VNR. $N^V$ represents the collection of virtual nodes in the VNR, and $n^v$ represents a certain virtual node. $L^V$ represents the virtual link set in the VNR, and $l^s$ represents one of the determined virtual links. $A_N^V$ represents the attribute set of virtual node. For a specific virtual node $n^v$, its attributes include CPU resource requirement $CPU(n^v)$, delay level requirement $DELAY(n^v)$ and security level requirement $SR(n^v)$. $A_L^V$ represents the attribute set of virtual link. For a specific virtual link $l^v$, its attributes include bandwidth resource requirement $BW(l^v)$ and delay level requirement $DELAY(l^v)$.

The undirected weighted graph $G^S=\{N^S,L^S,A_N^S,A_L^S\}$ is used to build the mathematical model for the substrate network. $G^S$ represents the entire substrate network. $N^S$ represents the node set of the substrate network, and $n^s$ represents a certain substrate node. $L^S$ represents the set of links in the substrate network, and $l^s$ represents a certain substrate link. $A_N^S$ represents the attribute set of substrate nodes. For a specific substrate node $n^s$, its attributes include available CPU resource $CPU(n^s)$, delay level $DELAY(n^s)$ and security level $SL(n^s)$. $A_L^S$ represents the attribute set of the substrate link. For a specific physical link $l^s$, its attributes include the available bandwidth resource $BW(l^s)$ and delay level $DELAY(l^s)$.

Fig. 1 shows the topology of a virtual network and a substrate network. The virtual network consists of two virtual nodes and one virtual link. The three numbers next to the virtual node represent the CPU resource demand, delay demand level and security demand level of the virtual node. The two numbers on the virtual link represent the bandwidth resource demand and delay demand level of the virtual link. For a substrate network, the three numbers next to each substrate node represent the current available CPU resources, delay level and security level of the substrate node. Two numbers on each substrate link represent the amount of bandwidth resources and delay level that the substrate link can provide.

\begin{figure}[!htp]
\centering
\includegraphics[width=0.7\columnwidth]{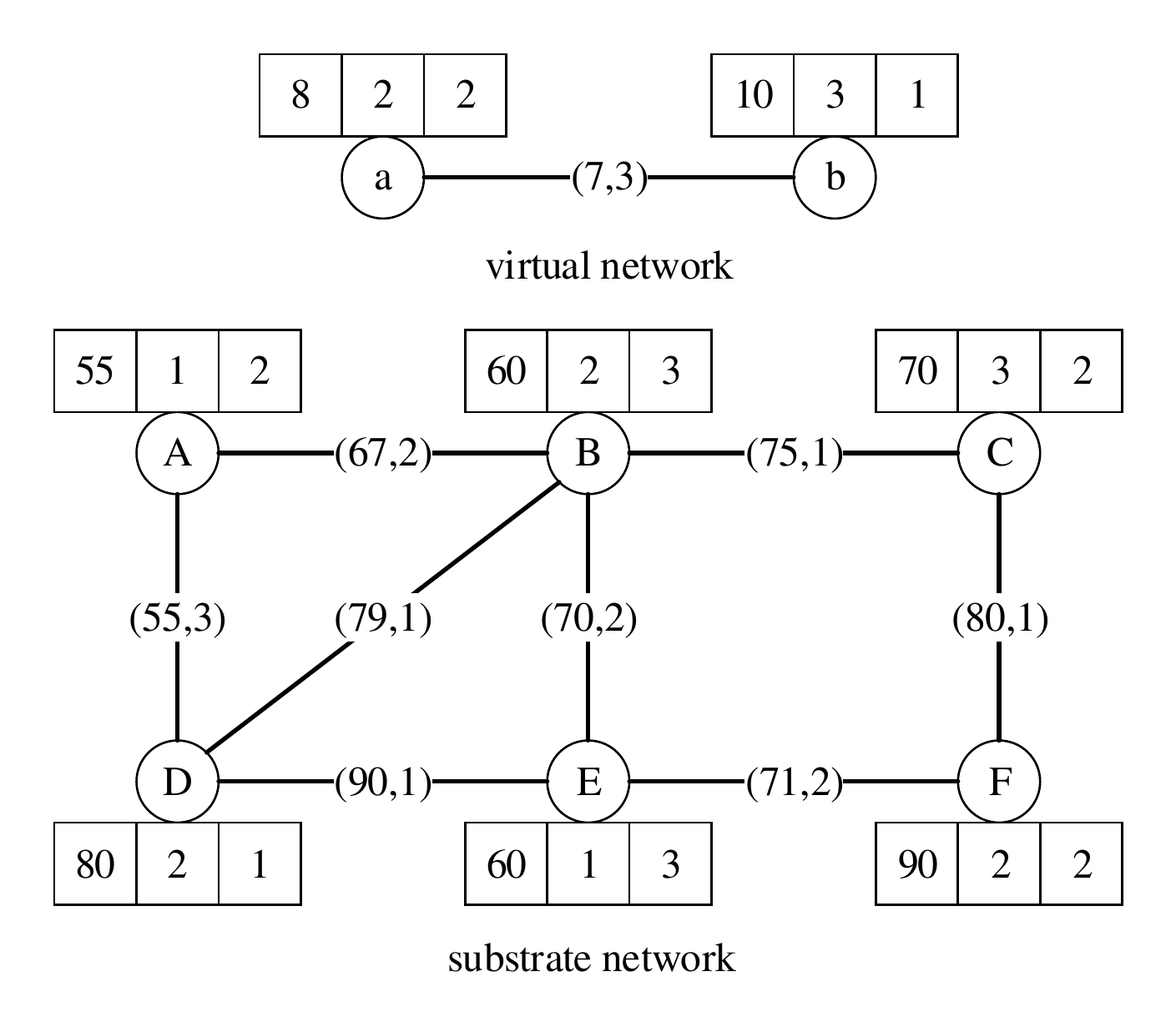}
\caption{Diagram of virtual network and substrate network.}
\label{fig_1}
\end{figure}

\subsection{Constraints}

The embedding of VNRs is limited by the number of underlying network resources. It is impossible for virtual network to be embedded on the underlying network without limitation. The successful embedding of every VNR consumes a certain amount of network resources, mainly including the CPU resources of nodes and the bandwidth resources of links. In addition, due to the need for collaborative consideration of differentiated QoS and security of VNE, we stipulate that virtual nodes and virtual links can only be mapped to substrate nodes and substrate links whose delay level is not greater than its delay demand level. At the same time, it is necessary to ensure that the virtual node maps to a substrate node whose security level is not less than its security requirement level. With the above constraints, we can ensure that the VNE meets the differentiated QoS requirements and security requirements of CPU, bandwidth and delay. We will be involved in the formulation of constraints.

The current available CPU resources of substrate node $n^s$ can be represented by the remaining CPU resources:

\begin{equation}
\begin{aligned}
CPU(n^s)=CPU_{initial}(n^s)-\sum_{all(n^v \uparrow n^s)}CPU(n^s).
\end{aligned}
\end{equation}

$CPU(n^s)$ represents the current remaining CPU resources of substrate node $n^s$. $CPU_{initial}(n^s)$ represents the initial total CPU resources of substrate node $n^s$. $\sum_{all(n^v \uparrow n^s )}CPU(n^s)$ represents the CPU resources consumed by all virtual nodes mapped to substrate node $n^s$. The symbol $n^v \uparrow n^s$ indicates that virtual node $n^v$ is mapped to substrate node $n^s$.

The currently available bandwidth resources of substrate link $l^s$ can be represented by the remaining bandwidth resources:

\begin{equation}
\begin{aligned}
BW(l^s)=BW_{initial}(l^s)-\sum_{all(l^v \uparrow l^s)}BW(n^s).
\end{aligned}
\end{equation}

$BW(l^s)$ represents the current remaining bandwidth resources of substrate link $l^s$. $BW_{initial}(l^s)$ represents the initial total bandwidth resources of substrate link $l^s$. $\sum_{all(l^v \uparrow l^s)}BW(n^s)$ represents the bandwidth resources consumed by all virtual links mapped to substrate link $l^s$. The symbol $l^v \uparrow l^s$ indicates that virtual link $l^v$ is mapped to physical link $l^s$.

\begin{equation}
\begin{aligned}
if\,\,n^v \uparrow n^s\,,CPU(n^v) \le CPU(n^s),
\end{aligned}
\end{equation}
\begin{equation}
\begin{aligned}
if\,\,l^v \uparrow l^s\,,BW(l^v) \le BW(l^s).
\end{aligned}
\end{equation}

Formula (3) and formula (4) respectively represent the CPU resource constraint and bandwidth resource constraint embedded in the virtual network.

\begin{equation}
\begin{aligned}
if\,\,n^v \uparrow n^s\,,DELAY(n^v) \ge DELAY(n^s),
\end{aligned}
\end{equation}
\begin{equation}
\begin{aligned}
if\,\,l^v \uparrow l^s\,,DELAY(l^v) \ge DELAY(l^s).
\end{aligned}
\end{equation}

Formula (5) and formula (6) respectively represent the node delay constraint and link delay constraint embedded in the virtual network. Virtual node $n^v$ can only be mapped to substrate node $n^s$ which is not greater than its delay requirement level. Virtual link $l^v$ can only be mapped to substrate link $l^s$ which is not greater than its delay requirement level.

\begin{equation}
\begin{aligned}
if\,\,n^v \uparrow n^s\,,SR(n^v) \le SL(n^s).
\end{aligned}
\end{equation}

Formula (7) represents the security constraints embedded in the virtual network. We set security requirement level for each virtual node and security level for each substrate node. Virtual node $n^v$ can only be mapped to substrate node $n^s$ which is not less than its security requirement level.

\subsection{Evaluating Indicators}

We take the long-term average revenue, acceptance rate and long-term revenue consumption ratio of VNE as the indexes to evaluate the performance of the algorithm. Because these three indicators can reflect the differentiated QoS requirements of CPU resource consumption, bandwidth resource consumption, delay constraint and security constraint to a certain extent.

The revenue of VNE is as follows:

\begin{equation}
\begin{aligned}
R(G^V,t)=\sum_{n^v \in N^V}CPU(n^v)+\sum_{l^v \in L^V}BW(l^v),
\end{aligned}
\end{equation}
where $R(G^V,t)$ represents the revenue embedded in the virtual network within the time period $t$ of the arrival of the VNR. $\sum_{n^v \in N^V}CPU(n^v)$ represents the CPU resource revenue obtained from the CPU resource consumed by virtual node $n^v$. $\sum_{l^v \in L^V}BW(l^v)$ represents the bandwidth resource revenue corresponding to the bandwidth resource consumed by the virtual link $l^v$. The revenue of VNE are determined by the sum of CPU resources and bandwidth resources consumed by VNRs.

The long-term average revenue of embedded virtual network is as follows:

\begin{equation}
\begin{aligned}
R=\lim_{T \to \infty}\frac{\sum_{t=0}^{T}R(G^V,t)}{T}.
\end{aligned}
\end{equation}

The consumption of VNE is expressed as follows:

\begin{equation}
\begin{aligned}
C(G^V,t)=\sum_{n^v \in N^V}CPU(n^v)+\sum_{l^v \in L^V}BW(l^v)\times hops(l^v),
\end{aligned}
\end{equation}
where $C(G^V,t)$ represents the consumption of VNE within the time period $t$ of VNR arrival. $\sum_{n^v \in N^V}CPU(n^v)$ represents the CPU resources consumed by virtual node $n^v$. $\sum_{l^v \in L^V}BW(l^v)\times hops(l^v)$ represents the bandwidth resources consumed by the virtual link $l^v$. Since a virtual link may be divided into multiple substrate links, $hops(l^v)$ represents the number of hops of the virtual link $l^v$. The consumption of VNE is determined by the sum of CPU resources and bandwidth resources consumed by VNRs.

The ratio of long-term revenue consumption embedded in virtual network is expressed as follows:

\begin{equation}
\begin{aligned}
R/C=\lim_{T \to \infty}\frac{\sum_{t=0}^{T}R(G^V,t)}{\sum_{t=0}^{T}C(G^V,t)}.
\end{aligned}
\end{equation}

The acceptance rate of VNE is expressed as follows:

\begin{equation}
\begin{aligned}
ACC=\lim_{T \to \infty}\frac{\sum_{t=0}^{T}num(VNR_{acc})}{\sum_{t=0}^{T}num(VNR_{arr})},
\end{aligned}
\end{equation}
where $\sum_{t=0}^{T}num(VNR_{acc})$ represents the number of VNRs successfully embedded in the time range $t$. $\sum_{t=0}^{T}num(VNR_{arr})$ represents the total number of VNRs that arrive in the time range $t$.

\subsection{An Example}

Fig. 2 shows the different situations of two kinds of VNE. In case 1, the node mapping relationship is $a \uparrow A$, $b \uparrow B$, and the virtual link does not have path segmentation at this time. In case 2, the node mapping relationship is $a \uparrow F$, $b \uparrow D$, and the virtual link has path splitting. In both cases, the successful embedding of VNRs consumes the corresponding CPU resources and bandwidth resources. The delay constraints and security constraints are also satisfied. But in both cases, the ratio of revenue to consumption is different.

\begin{figure}[!htp]
\centering
\includegraphics[width=0.7\columnwidth]{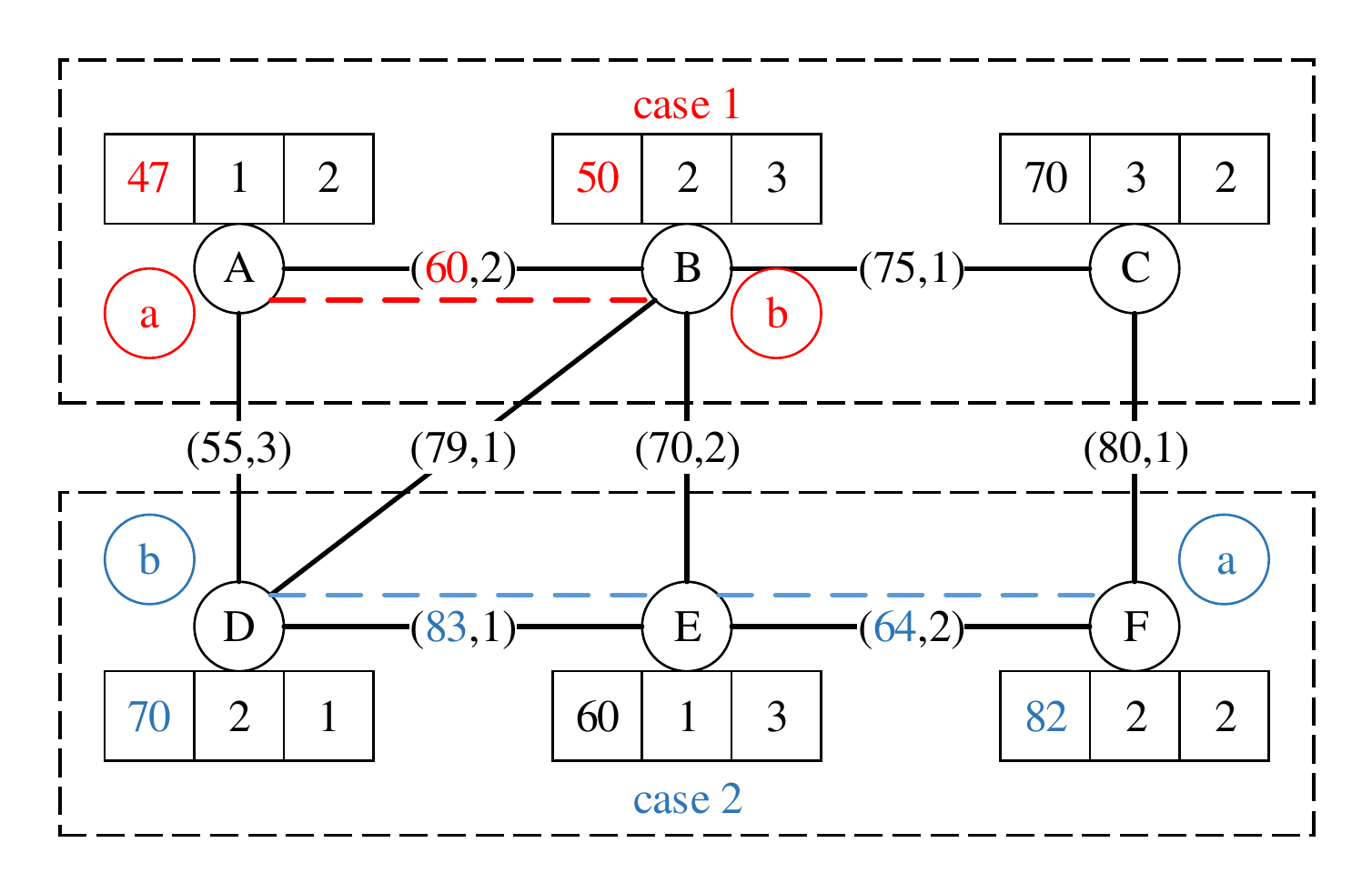}
\caption{Example of virtual network embedding.}
\label{fig_2}
\end{figure}

In case 1, the revenue of VNE is:

\begin{equation}
\begin{aligned}
R(G^V,t)&=\sum_{n^v \in N^V}CPU(n^v)+\sum_{l^v \in L^V}BW(l^v) \\
&=CPU(n_a^v)+CPU(n_b^v)+BW(l_{ab}^v) \\
&=8+10+7 \\
&=25
\end{aligned}
\end{equation}

The consumption of VNE is as follows:

\begin{equation}
\begin{aligned}
C(G^V,t)&=\sum_{n^v \in N^V}CPU(n^v)+\sum_{l^v \in L^V}BW(l^v)\times hops(l^v)\\
&=CPU(n_a^v)+CPU(n_b^v)+BW(l_{ab}^v) \times hops(l_{ab}^v)\\
&=8+10+7 \times 1\\
&=25
\end{aligned}
\end{equation}

So the ratio of revenue to consumption in case 1 is 1.

In case 2, the revenue of VNE is:

\begin{equation}
\begin{aligned}
R(G^V,t)&=\sum_{n^v \in N^V}CPU(n^v)+\sum_{l^v \in L^V}BW(l^v) \\
&=CPU(n_a^v)+CPU(n_b^v)+BW(l_{ab}^v) \\
&=8+10+7 \\
&=25
\end{aligned}
\end{equation}

The consumption of VNE is as follows:

\begin{equation}
\begin{aligned}
C(G^V,t)&=\sum_{n^v \in N^V}CPU(n^v)+\sum_{l^v \in L^V}BW(l^v)\times hops(l^v)\\
&=CPU(n_a^v)+CPU(n_b^v)+BW(l_{ab}^v) \times hops(l_{ab}^v)\\
&=8+10+7 \times 2\\
&=32
\end{aligned}
\end{equation}

So the ratio of revenue to consumption in case 2 is $\frac{25}{32}$.

It can be seen that path segmentation will result in greater cost of resource consumption. Therefore, in the design of VNE algorithm, it should try to avoid the generation of path segmentation.

\section{Implementation of VNE algorithm Based on Differentiated QoS and Security Requirements}

\subsection{The Framework of VNE algorithm Based on DRL}

\begin{figure}[!htp]
\includegraphics[width=1.0\columnwidth]{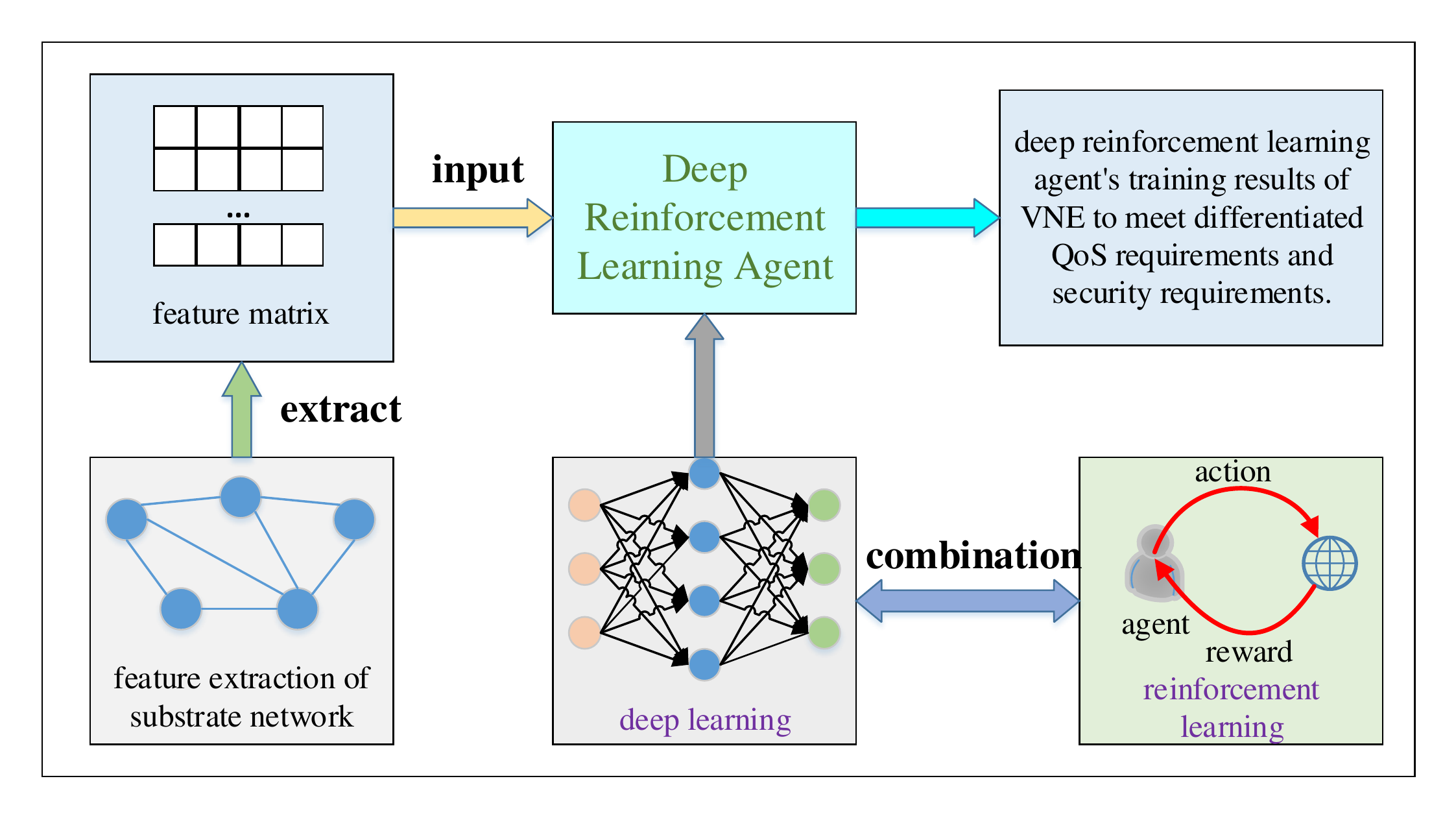}
\caption{The framework of VNE based on DRL.}
\label{fig_3}
\end{figure}

With the continuous improvement of intelligent learning algorithm performance and the expansion of its application range, the application of DRL method to VNE algorithm will become the mainstream to solve the problem of VNE. The key of using DRL method to solve the problem of VNE lies in which kind of neural network is used to train the agent and how to create a realistic network environment for the agent. The embedded algorithm framework of virtual network based on DRL is shown in Fig. \ref{fig_3}.

In order to train the DRL agent in a more real network environment, we extract four important attributes of the substrate network according to the differentiated QoS requirements and security requirements, so as to build the agent training environment. Feature extraction is described in detail below. DRL combines the advantages of deep learning and reinforcement learning, and agents can be well trained in our own policy network. Finally, a good result of VNE can be obtained.

\subsection{Network Feature Extraction}

The purpose of feature extraction of substrate network is to create a more real network environment for DRL agent. The DRL agent can make the best decision only when it is trained in the substrate network environment. Under the NV architecture, the substrate network is very complex and the network features available for extraction are very rich. Considering the computing power of our own policy network, if we extract too many network features, it will cause great computational complexity and reduce the performance of the algorithm. For differentiated QoS requirements and security requirements, we extract the following four attributes for each physical node:

\textit{(1)} CPU resources. CPU resource is one of the most important resources in the network environment, which is an important factor affecting the cost of VNE. The CPU resource calculation method of the substrate node is shown in formula (1).

\textit{(2)} The sum of bandwidth. The bandwidth sum of all links connected to a substrate node, which can reflect the bandwidth resource requirements of users. The link bandwidth sum connected to the substrate node is expressed as:
\begin{equation}
\begin{aligned}
SUM\_BW(n^s)=\sum_{l^s \in L_{n^s}^S}BW(l^s).
\end{aligned}
\end{equation}

\textit{(3)} Delay. We set the delay attribute for both the substrate node and the substrate link. Virtual nodes and links can only be mapped to substrate nodes and links that are no higher than their delay requirements \cite{ggg}. The specific expression is shown in formula (5).

\textit{(4)} Safety level. We set the security level for each substrate node. Virtual nodes can only be mapped to substrate nodes that are no lower than the security requirement level \cite{gg}. The specific expression is shown in formula (7).

CPU resources and bandwidth resources are the main resources consumed by VNRs. The evaluation criteria for evaluating the performance of VNE algorithm are also designed based on these two attributes. In order to meet the differentiated QoS requirements of network end-users/applications, it is necessary to consider the bandwidth resources and delay together. They deal with different QoS scenarios respectively. In addition, in view of the security problems in the network environment, innovatively considering the security attribute in the VNE problem can provide a new idea for the security of the VNE algorithm.

The extracted node attributes are concatenated into a four-dimensional feature vector. The feature vector of substrate node $n^s$ is expressed as:

\begin{equation}
\begin{aligned}
v^{n^s}=\{CPU(n^s),BW(n^s),DELAY(n^s),RL(n^s)\}.
\end{aligned}
\end{equation}

By combining the feature vectors of all substrate nodes, a four-dimensional feature matrix can be obtained. As the input of the policy network, the DRL agent is trained in the environment of the feature matrix. The feature matrix is expressed as:

\begin{equation}
\begin{aligned}
M_f=\{v_1^{n^s},v_2^{n^s},...,v_k^{n^s}\}.
\end{aligned}
\end{equation}

\subsection{Policy Network Construction}

\begin{figure}[!htp]
\includegraphics[width=1.0\columnwidth]{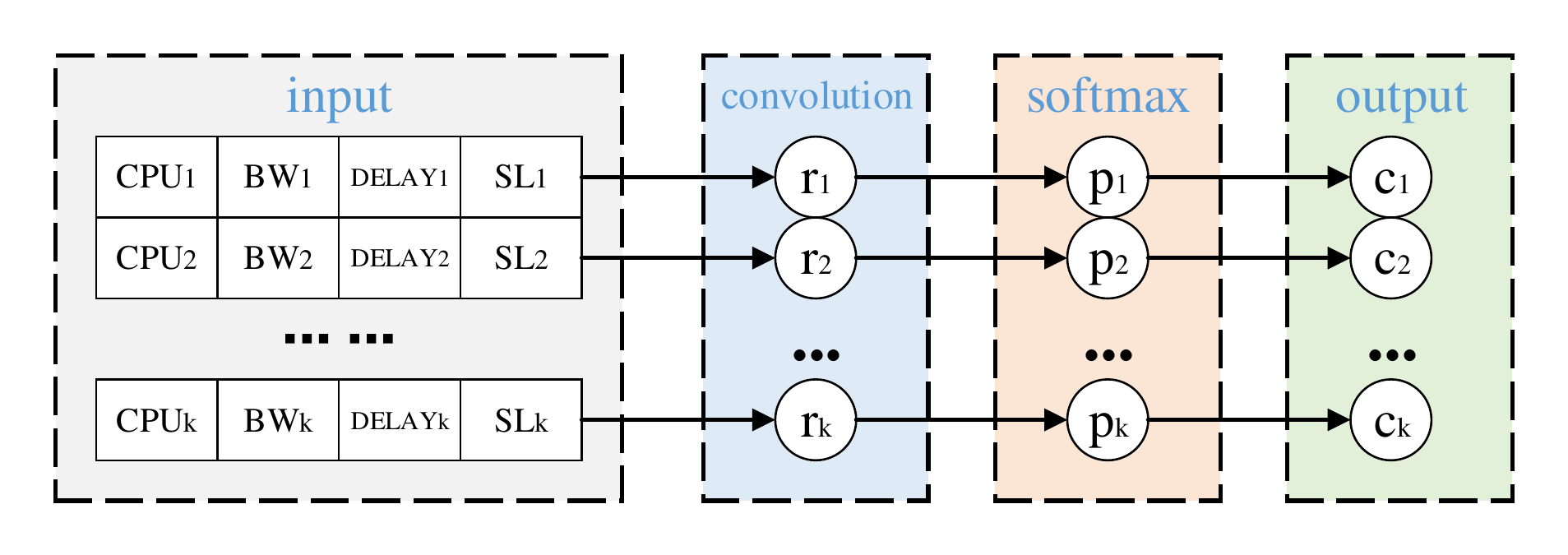}
\caption{Policy network.}
\label{fig_4}
\end{figure}

We use the basic elements of artificial neural network to build a simple policy network as a DRL agent. The extracted feature matrix is used as the input of the policy network, and the agent learns and trains in this environment, in order to deduce the probability of each substrate node being mapped. The resulting policy network is shown in Fig. \ref{fig_4}. It mainly includes input layer, convolution layer, softmax layer and output layer.

The input layer is used to receive the feature matrix extracted from the substrate network, then use the policy network to evaluate the node attributes in the feature matrix. Convolution operation is performed on the feature vectors in the convolution layer to obtain the available resource vector $r_i$ of each eigenvector. The operation method is as follows:

\begin{equation}
\begin{aligned}
r_i=\omega \cdot v_i+o,
\end{aligned}
\end{equation}
where $r_i$ is the available resource vector of the ith feature vector. $\omega$ is the convolution kernel weight vector of the convolution layer. $v_i$ is the ith feature vector and $o$ is the offset.

In the softmax layer, a probability is generated for each node according to the available resource vector of each node, and the substrate nodes are ordered according to this probability. Softmax function is the extension of logical regression. It can convert n-dimensional vector into real value between 0 and 1 \cite{a5}. The calculation method is as follows:

\begin{equation}
\begin{aligned}
p_i=\frac{e^{r_i}}{\sum_ke^{r_k}},
\end{aligned}
\end{equation}
where $p_i$ represents the mapping probability of the ith substrate node. $k$ is the total number of feature vectors.

Finally, a set of available substrate nodes and their mapping probabilities are output at the output layer.

\subsection{Training and Testing}

In the training phase, whenever a VNR arrives, the policy network will extract a feature matrix from the substrate network as the input, so the feature matrix is dynamic. According to the input feature matrix, the agent can learn the most real situation of the substrate network and make the optimal decision. In reinforcement learning, in order to encourage the agent to make the best decision, a reward signal is usually set for the agent. The agent will decide which action to take according to the size of the reward signal. The size of the reward signal also depends on whether the agent's action is beneficial to it. The two are interactive. If an action taken by an agent receives a larger reward, the agent will take similar actions to accumulate the reward. In the problem of VNE, the revenue consumption ratio of VNE is usually used as the reward signal of agent. On the one hand, as an important evaluation index of VNE, the index has certain representativeness. On the other hand, the index reflects the utilization rate of the underlying network resources to a certain extent. There is a positive correlation between revenue consumption ratio and reward signal.

According to the method of supervised learning, a manual label is introduced for each feature vector in the policy network. Suppose that the manual label is introduced for the ith feature vector, then the label is 0 except that the ith position is 1. That is:

\begin{equation}
\begin{aligned}
label_i=\{0,0,...,1,...0\}.
\end{aligned}
\end{equation}

The cross entropy loss is calculated as follows:

\begin{equation}
\begin{aligned}
Loss(label,c)=-\sum_ilabel_i \log(c_i),
\end{aligned}
\end{equation}
where $label_i$ and $c_i$ are the ith element of label and the output of policy network respectively.

Using back propagation to calculate the gradient of parameters in the policy network:

\begin{equation}
\begin{aligned}
g=\alpha \cdot r \cdot g_s,
\end{aligned}
\end{equation}
where $\alpha$ is the size of training gradient. $r$ is the size of reward signal. $g_s$ is the stacking gradient.

The training process of DRL agent is shown in algorithm 1.

\begin{algorithm}
  \caption{Training algorithm}
  \begin{algorithmic}[1]
    \Require
        {$epoch$;\,$\alpha$;\,$trainingset$};
    \Ensure
        {$trained\,\,parameters\,\,in\,\,policy\,\,network$};
    \State $initialize\,\,parameters\,\,in\,\,policy\,\,network$;
    \While {$iteration\,<\,epoch$}
    \For {$request \in trainingset$}
    \State $counter=0$;
    \For {$node \in request$}
    \State $M_f=getFeatureMatrix()$;
    \State $p=policyNetwork.output(M_f)$;
    \State $host=sample(p)$;
    \State $gradient(host)$;
    \EndFor
    \If {$sn.cpu \geq vn.cpu\,and\,sl \geq sr\,and\,vn.delay \geq sn.delay$}
    \If {$isMapped(\forall\,node \in request,\,\forall\,link \in request)$}
    \State $reward=revToCost(request)$;
    \State $multiplyGradient(reward,\alpha)$;
    \Else
    \State $claer gradients$;
    \EndIf
    \EndIf
    \State $counter++$;
    \If {$counter == batch\_size$}
    \State {$counter = 0$};
    \EndIf
    \EndFor
    \State {$iteration++$};
    \EndWhile
    \State $return\,parameters$;
  \end{algorithmic}
\end{algorithm}

In the test phase, according to the node mapping probability obtained in the training phase, the virtual nodes are mapped in turn. Finally, the BFS is used to map the virtual links. The test process is shown in algorithm 2.

\begin{algorithm}
  \caption{Test algorithm}
  \begin{algorithmic}[1]
     \Require
        {$test\_set$};
    \Ensure
        {$long\,\,term\,\,average\,\,revenue$},
        {$long\,\,term\,\,revenue\,\,consumption\,\,ratio$},
        {$VNR\,\,acceptance\,\,rate$};
    \State {$initialize\,\,paraqmeters\,\,in\,\,policy\,\,network$};
    \For {$request \in test\_set$}
    \For {$node \in request$}
    \State $M_f=getFeatureMatrix()$;
    \State $host=maxProbability(p)$;
    \EndFor
    \State $BFSLinkMap(request)$;
    \If {$isMapped(\forall\,node \in request, \forall\, link \in request)$}
    \State $return\,(success)$;
    \EndIf
    \EndFor
  \end{algorithmic}
\end{algorithm}

\section{Experimental Setup and Result Analysis}

In this part, we first introduce the setting of the experimental environment, then we will show the experimental results and analyze them.

\subsection{Experimental Setup}

We build a medium scale substrate network with 100 substrate nodes and 570 substrate links. The initial CPU resources of each substrate node are evenly distributed between 50 units and 100 units. In order to reflect the different requirements of users for delay characteristics, we set the delay level for each substrate node. The delay level is evenly distributed between 1 and 3. In order to reflect the user's demand for network security, we set the security level for each physical node. The security level is evenly distributed between 1 and 3. The initial bandwidth resources of each substrate link are evenly distributed between 50 units and 100 units, and the delay level is evenly distributed between 1 and 3.

We generated 2000 VNRs, of which the first 1000 were used as training sets and the last 1000 as test sets. Each VNR contains 2 to 10 different nodes, each node has a 50\% probability of interconnection. The CPU resource demand of each virtual node is evenly distributed from 1 unit to 50 units. In order to reflect the different requirements of users for delay characteristics, we set the delay requirement level for each substrate node. The delay demand level is evenly distributed from 1 to 3. In order to reflect the user's demand for network security, we set the security requirement level for each virtual node. The security demand level is evenly distributed from 1 to 3. The bandwidth resource demand of each virtual link is evenly distributed between 1 unit and 50 units, and the delay demand level is evenly distributed between 1 and 3.

The arrival process of VNRs is simulated by Poisson distribution. The average arrival time of every 100 time units reaches 4 VNRs and the duration of each request follows the exponential distribution. We trained 100 epoch agents with gradient descent method, and the learning rate was set to 0.005.

The above experimental data are summarized in Table 1.

\begin{table}
\centering
\caption{Parameter setting}
\renewcommand\arraystretch{1.5}
\begin{tabular}{p{40mm}|p{30mm}}
\hline
Parameter name & Parameter value  \\
\hline
number of substrate nodes & 100 \\
number of substrate links & 570 \\
initial CPU resources & U[50, 100] \\
delay level of nodes & U[1, 3] \\
safety level of nodes & U[1, 3] \\
initial bandwidth resources & U[50, 100] \\
delay level of links & U[1, 3] \\
\hline
number of nodes per VNR & U[2, 10]  \\
node connection rate & 50\%  \\
CPU resource demand & U[0, 50] \\
delay demand level of nodes & U[1,3] \\
safety demand level of nodes & U[1,3] \\
bandwidth resource demand & U[0, 50] \\
delay demand level of links & U[1,3] \\
\hline
\end{tabular}
\end{table}

\subsection{Results and Analysis}

\subsubsection{Training Results and Analysis}

\begin{figure*}[!htp]
\centering
\includegraphics[width=1.0\textwidth]{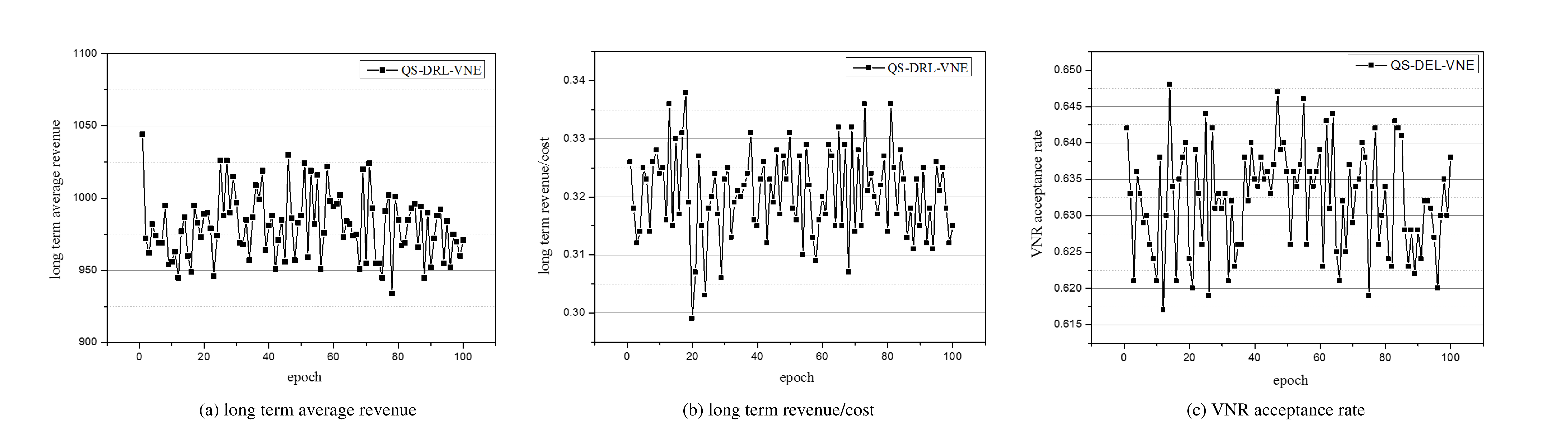}
\caption{Results on training set.}
\label{fig_5}
\end{figure*}

Because the problem of VNE has been proved to be NP-hard and uncertain, the ultimate goal is to find an optimal solution. In order to achieve good results, we put the DRL agent on 100 epoch training sets. An epoch refers to the process of sending all the data into the network to complete a complete training. Through training, we get the stability degree of agent in three indexes: long term average revenue, long term revenue consumption ratio and VNR acceptance rate. The results are shown in Fig. \ref{fig_5}.

In the initial stage of training, since the parameters of the policy network are randomly initialized, the agent will randomly take actions to explore the possibility of achieving good results, and the stability is poor at this time. In the middle of the training stage, as the agent becomes more and more familiar with the network environment, the agent will continue to find good solutions. At this time, the agent will get a larger reward signal. Meanwhile, the stability of the agent begins to improve gradually. In the later stage of training, the learning ability of the agent is limited by the performance of the policy network. At this time, the agent accumulates a certain degree of rewards, and the actions taken gradually tend to be stable, so the fluctuation range is small. The gradual stability of the curve proves the effectiveness of agent training, which lays a good foundation for the application of DRL agent in test set.

\begin{figure}[!htp]
\centering
\includegraphics[width=0.7\columnwidth]{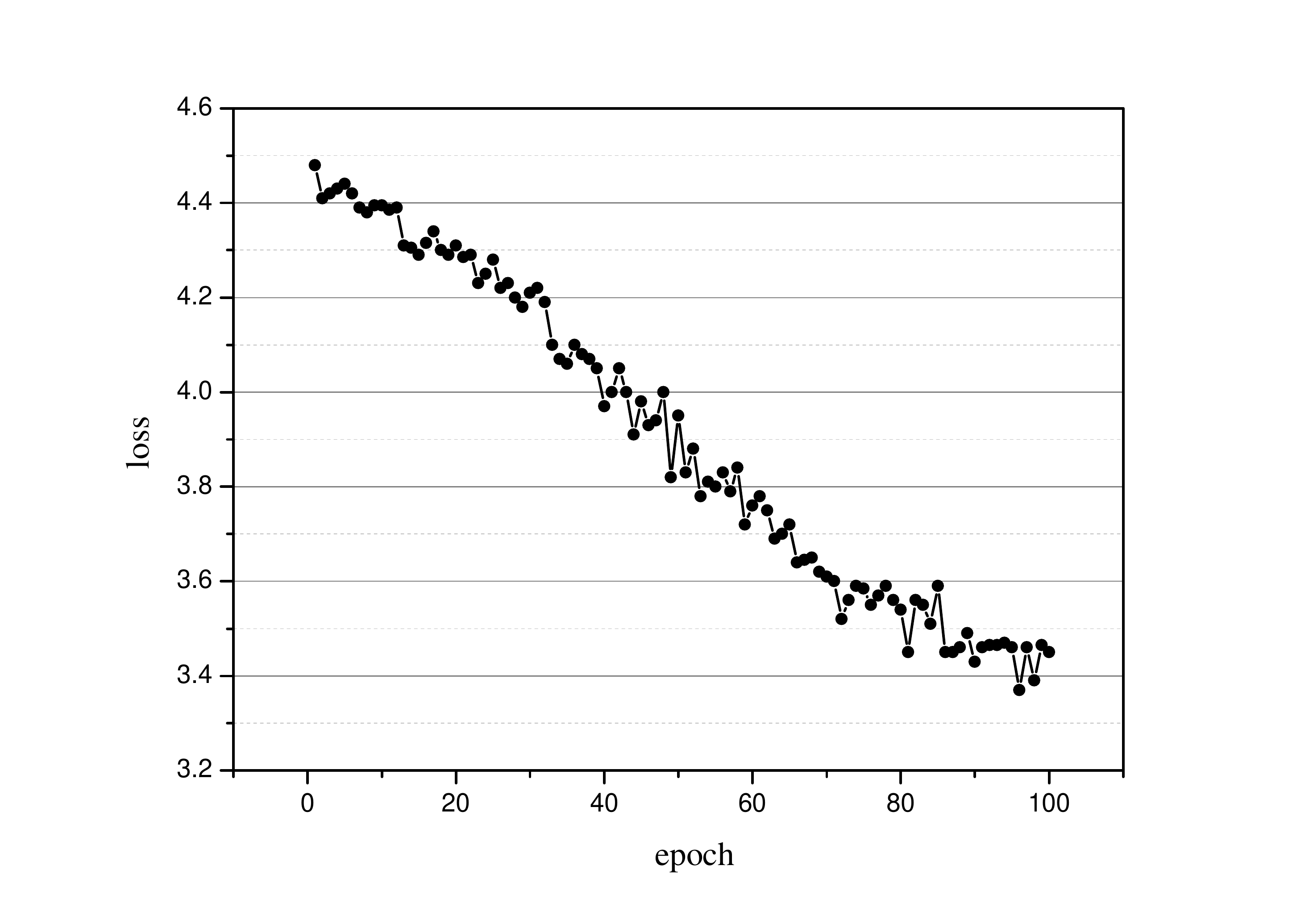}
\caption{Cross entropy loss.}
\label{fig_6}
\end{figure}

Fig. \ref{fig_6} shows the change of cross entropy loss during the training phase. It can be seen from the figure that the loss value of training keeps decreasing, which also proves that the training of policy network is effective.

It can be seen from the figure that the long term average revenue embedded in the virtual network and the acceptance rate of VNRs are decreasing with the increase of time, because these two indicators are limited by the number of network resources. When the underlying network resources are consumed, the number of VNRs it can carry will continue to decrease. Therefore, the long term average revenue and acceptance rate will show a downward trend. The change of the revenue consumption ratio embedded in the virtual network has nothing to do with the quantity of the underlying network resources, so the index does not show a downward trend.

\subsubsection{Test Results and Analysis}

After the training, the DRL agent is put in the test set to test, so as to prove the effectiveness of the algorithm. Since the mapping probability of each substrate node is obtained in the training phase, the mapping of virtual nodes is directly based on the probability in the test phase.

\begin{figure*}[!htp]
\centering
\includegraphics[width=1.0\textwidth]{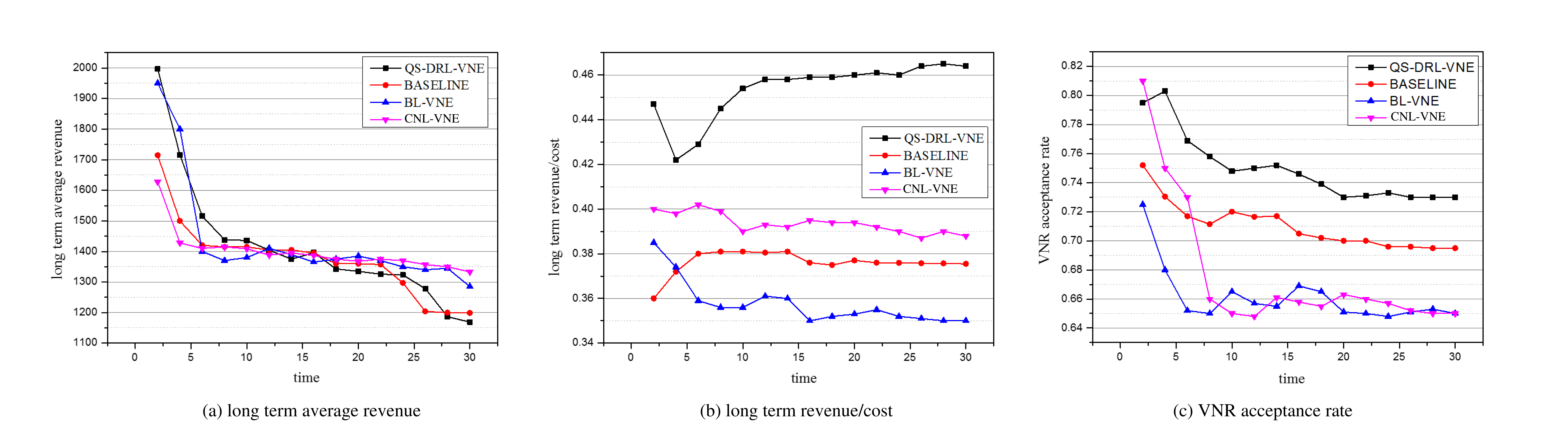}
\caption{Results on test set.}
\label{fig_7}
\end{figure*}

We compare the DRL-VNE algorithm based on differentiated QoS and security requirements (QS-DRL-VNE) with BASELINE algorithm \cite{c9}, BL-VNE algorithm \cite{c6} and CNL-VNE algorithm \cite{c10}. BASELINE algorithm is a typical VNE algorithm based on intelligent learning method. BL-VNE algorithm is based on the mapping cost of virtual network. CNL-VNE algorithm is a security VNE algorithm. The core ideas of several algorithms are listed in Table 2. The experimental results show the performance of the algorithm in three aspects: long term average revenue, long term revenue consumption ratio and VNR acceptance rate, as shown in Fig. \ref{fig_7}.

\begin{table}
\centering
\caption{Parameter setting}
\renewcommand\arraystretch{1.5}
\begin{tabular}{|p{20mm}|p{100mm}|}
\hline
Algorithm & Content \\
\hline
QS-DRL-VNE & Focus on CPU, bandwidth, delay and security differentiated QoS algorithm. By extracting the features of the substrate network to form the feature matrix, the DRL agent is trained in this environment, and finally the node mapping probability is deduced. Use the BFS to complete the link mapping.  \\
\hline
BASELINE & The formula $H(n^s)=CPU(n^s)\sum_{l^s \in L(n^S)}BW(L^S)$ is used to sort the substrate nodes, and the BFS is used to complete the link mapping. \\
\hline
BL-VNE & Using greedy strategy to complete node mapping, using shortest path algorithm to complete link mapping heuristic VNE algorithm. \\
\hline
CNL-VNE & VNE algorithm based on D-ViNE. In the process of node mapping, security constraints are added and the revenue and costs of VNE are modified according to security constraints. \\
\hline
\end{tabular}
\end{table}

According to the comparison results of the three indexes, the DRL-VNE algorithm based on differentiated QoS and security requirements performs better than the other three algorithms, for two main reasons. First, based on differentiated QoS and security requirements, the DRL-VNE algorithm adopts efficient intelligent learning method to serve the problem of VNE, and efficiently solves the decision-making and optimization process of VNE. Compared with BL-VNE algorithm and CNL-VNE algorithm, intelligent learning algorithm has more advantages. Secondly, the DRL-VNE algorithm based on differentiated QoS and security requirements reasonably extracts the substrate network features, so that the DRL agent can be trained in a more real network environment, and finally achieves better results in the test set compared with the BASELINE algorithm. The above results show that the DRL-VNE algorithm based on differentiated QoS and security requirements is effective.

\section{Conclusion}

This paper combines the DRL algorithm with the VNE algorithm, and creatively solves the differentiated QoS and security requirements of network end users/ applications. In the framework of NV, the QoS and security requirements of network end users/applications are ultimately the problem of VNE. Using DRL method can improve the decision-making and optimization ability of VNE algorithm.

We attribute the QoS and security requirements of network end users/applications to four network indicators: CPU, bandwidth, delay and security. The DRL agent is trained in the network environment composed of these four attributes and finally the mapping probability of each substrate node is obtained. The experimental results show that it is feasible to solve the problem of VNE by this method and good results have been achieved. Therefore, it is of great practical significance to apply the algorithm to solve the differentiated QoS and security requirements of network end users/applications.

\begin{acknowledgements}
This work is partially supported by the Major Scientific and Technological Projects of CNPC under Grant ZD2019-183-006, partially supported by the Project ``Research on multi domain virtual network mapping algorithm for differentiated QoS requirements" supported by Shandong Provincial Natural Science Foundation, and partially supported by ``the Fundamental Research Funds for the Central Universities" of China University of Petroleum (East China) under Grant 20CX05017A.
\end{acknowledgements}

\end{document}